\documentstyle[preprint,aps]{revtex}

\begin{document}

\draft

\title{Fractional Vortices as Evidence of Time-Reversal
	Symmetry Breaking in High-Temperature Superconductors}

\author{M. Sigrist\cite{sigrist}}
\address{Department of Physics, Massachussets Institute of Technology,
	Cambridge, MA 02139}

\author{D. B. Bailey\cite{bailey}}
\address{Department of Physics, Stanford University, Stanford, CA 94305}

\author{and \\ R. B. Laughlin\cite{laughlin}}
\address{Department of Physics, Stanford University, Stanford, CA 94305\\
	and \\
	Lawrence Livermore National Laboratory,
	P. O. Box 808, Livermore, CA 94550}

\maketitle

\begin{abstract}
We argue that recent experiments by Kirtley {\it et al.}\cite{KIR} may show
evidence of time reversal symmetry breaking in YBa$_2$Cu$_3$O$_7$ at
crystal grain boundaries.
We illustrate this through a Ginzburg-Landau model calculation.
Further experimental tests are proposed.
\end{abstract}


\narrowtext
In a recent paper, Kirtley {\it et al.}\cite{KIR} reported the
observation of magnetic defects at artificially engineered
grain boundaries in thin films of the high temperature superconductor
YBa$_2$Cu$_3$O$_7$ (YBCO).
The grain boundaries were the borders between a triangular
YBCO inclusion in a film of YBCO with the crystal axes misoriented
with respect to one another in the two domains (inside and outside
of the triangle). While the resolution of the magnetic microscope
($ \sim 10 \mu$m  which is roughly ten times the estimated Josephson
penetration depth $\lambda_J$\cite{AJM})
used for detection is
not sufficient to tell with absolute certainty, the observed magnetic
defects appear from their
shape and localization to be superconducting vortices carrying small
fractions of a flux quantum $\Phi_0=hc/2e$. These vortices are
attached mainly to the corners of the triangle, but occasionally
appear along the edges of the triangular inclusion.
 The purpose of this
Letter is to point out that the identification of these defects
as fractional vortices, if correct,
demonstrates that the materials in question have superconducting
order parameters, and thus ground states, that violate time
reversal symmetry.  The experiments cannot tell whether the
${\cal T}$-violation is a bulk or interface (grain boundary) effect.
Our argument is simply that the flux carried by a vortex measures the {\em
phase defect} of the order parameter along a closed path encircling
the vortex, and can therefore be fractional only if the order parameter
undergoes a phase jump $\Delta\phi$ {\em not} a multiple of $2\pi$
along this path.  The recent Josephson tunneling experiment of
Wollman {\em et al.}\cite{VAN} and the observation of half-integer
flux quanta by Tsuei {\em et al.}\cite{TSU} are specific examples of this for
which $\Delta\phi$ is an odd multiple of $\pi$. Because of specific
symmetry properties of the Josephson junctions their results were
interpreted as strong evidence for $ d_{x^2-y^2} $-wave pairing symmetry
in YBCO, which is a $ {\cal T} $-conserving superconducting state.
On the other hand, the recent experiment of
Kirtley and co-workers\cite{KIR} can be explained only if
$\Delta\phi$ is {\em not} a multiple of $\pi$, which requires
${\cal T}$-violation.

Let us briefly review the historical context of ${\cal T}$-violation in
unconventional (both heavy-fermion and high-$T_c$) superconductivity.
It has long been suspected that time reversal symmetry breaking is
responsible for some of the unusual magnetic properties of heavy
fermion superconductors, in particular (U,Th)Be$_{13}$ and UPt$_3$
\cite{GOR,LUK}.
The possible appearance of fractional vortices in these materials
has already been suggested \cite{VOL} and investigated theoretically
\cite{SRU}. The conditions under which a superconductor with a real order
parameter in the bulk phase may spontaneously break time reversal
symmetry have also been studied in the context of Ginzburg-Landau
theory\cite{PMS,SOU}.
Surfaces and domain walls were found under certain
conditions to favor the formation of a locally ${\cal T}$-violating state as
a means of lowering the energy cost of an inhomogeneous order
parameter\cite{SOU}.  ${\cal T}$-violation (specifically, a $d_{x^2-y^2}
+i\epsilon d_{xy}$ order parameter)
has been predicted in high-$T_c$ superconductivity via the anyon
technique applied to the $t$-$J$ model\cite{UNP,ROH}.  A superconducting
state with $s+id_{xy}$ symmetry has also been proposed\cite{KOT}.
However, none of the telltale signs of
${\cal T}$-violation has been detected in bulk measurements\cite{TAY}.
This does not preclude the existence of a complex order parameter at
surfaces and grain boundaries since bulk measurements are not
sensitive to the existence of such a phase. It is an additional point of
this Letter to show at least on a phenomenological level that such
states are indeed possible.

To illustrate our idea we first analyze the properties of
superconducting
states near an interface by means of a Ginzburg-Landau (GL) theory of
two complex order parameters, $ \eta_1 $ and $ \eta_2 $. These two
order parameters belong to pairing states of different symmetry,
e.g., $ d_{x^2-y^2} $ and $ d_{xy} $: $ \psi_1({\bf k}) = k^2_x - k^2_y $
and $ \psi_2({\bf k}) = k_x k_y $, which are non-degenerate under the
tetragonal ($D_{4h}$) crystal field symmetry assumed here. From this
we can derive a GL free energy functional of $ \eta_1 $ and
$ \eta_2 $ with the requirement that it be a scalar under all symmetries
of the system (for a review see
Ref.\cite{GOR}):
$ F = F_1 + F_2 + F_{12} $ with

\begin{equation}
F_i = \int d^3x\,[\alpha_i(T)|\eta_i|^2 + \beta_i|\eta_i|^4
	+K_i|{\bf D}\eta_i|^2]
\end{equation}

\begin{equation}
F_{12} = \int d^3x\,[\gamma|\eta_1|^2|\eta_2|^2+\delta(\eta_1^{*2}
	\eta_2^2+\eta_1^2\eta_2^{*2})]
\end{equation}

\noindent
where $\alpha_i(T) \propto T - T_{ci} $ ($ T_{ci} $, the bare bulk
transition temperature of the order parameter $ \eta_i $) and
$\beta_i,K_i,\gamma$ and $\delta$ are real
phenomenological parameters
which contain all the relevant physical information of microscopic
origin. The gradient terms are given
in the gauge invariant form $ {\bf D} = \nabla - i2 \pi {\bf A}/\Phi_0 $
with $ {\bf A} $ the vector potential ($ {\bf B} = \nabla
\times {\bf A} $) and $ \Phi_0 $ the flux quantum $ hc/2e $.  We note
that this is {\em not} the most general Ginzburg-Landau free energy
allowed by symmetry. For simplicity we include only terms
which are relevant for our discussion and, in particular, use an
isotropic gradient term. The boundary conditions at the interface
(surface or grain boundary) can be formulated in the standard way by

\begin{equation}
{\bf n} \cdot {\bf D} \eta_i = \eta_i/b_i, \hskip 0.7 cm i=1,2
\end{equation}

\noindent
at the interface. Here $ {\bf n} $ denotes the normal vector of the
interface and $ b_i $ is the so-called extrapolation length depending
on the properties and orientation of the interface (see \cite{GEN,GOR}).
Non-s-wave order parameters are often distorted in this manner in the
vicinity of an interface due to scattering effects \cite{GOR}.

Let us now examine the possibility
of $ {\cal T} $-violation at the interface. We shall assume,
as suggested from various experimental observations, that in the
bulk only the single component $ \eta_1=u_1e^{i\phi_1} $ exists,
while $ \eta_2=u_2 e^{i\phi_2} $ vanishes for all temperatures.  Thus,
we require that

\begin{equation}
u_1 = \tilde{u} (T)= \sqrt{-\alpha_1/2\beta_1},\qquad u_2 = 0
\end{equation}

\noindent
which is satisfied for all temperatures below $ T_{c1} $ under
the conditions

\begin{equation} \begin{array}{l}
(\gamma+2\delta {\rm cos} (2 \theta)) \tilde{u}^2-\alpha_2 > 0 \\ \\
T_{c_1} > T_{c_2} \end{array}
\end{equation}

\noindent
where $ \tilde{u} $ is the asymptotic value of $u_1$ in the bulk region
and $ \theta= \phi_1 - \phi_2 $ denotes the relative phase between
the two order parameter components.
With the choice $\delta>0$, the state with
$\theta=\pm\pi/2$ ($\psi({\bf k})=u\psi_1({\bf k})
\pm iv\psi_2({\bf k}$, or $d_{x^2-y^2} \pm i\epsilon d_{xy}$) is closest to
the instability although under condition, Eq.(5),
not stable for any temperature.
Thus, we may first treat
$u_1({\bf x})$ near a planar interface as though $ u_2$ were zero,
considering the GL differential equation obtained by variation of $F$ with
respect to $ \eta_1 $ (we neglect the vector potential for this discussion).
Note that $ u_1 $ depends only on the coordinate parallel to the normal
vector $ {\bf n} $ which we may choose to be parallel to the $ x $-axis with
the interface located at $ x=0 $.

\begin{equation}
K_1\frac{\partial^2 u_1}{\partial x^2} =
\alpha_1 u_1 +2\beta_1 u_1^3
\end{equation}

\noindent
As a solution to this equation we obtain

\begin{equation}
u_1(x) = \tilde{u} \tanh \left(\frac{x+x_0}{\xi} \right)
\end{equation}

\noindent
with $ \xi= \sqrt{2 K_1/\alpha_1} $ and $ x_0 = (\xi/2) \sinh^{-1}(4b_1/\xi) $.
Next, we ask whether this interface state could be unstable
against the admixture of a small component $ u_2 $. This question
can be answered
by analyzing the linearized GL-equation of $ u_2 $ for fixed $ u_1(x) $.

\begin{equation}
K_2\frac{\partial^2 u_2}{\partial x^2} =
\alpha_2 u_2 + (\gamma+2\delta \cos(2\theta)) u_1^2(x) u_2
\end{equation}

\noindent
It is easy to see that this equation
has the form of a Schr\"odinger equation for the wave function
$ u_2 $ of a particle in a potential well (including the boundary condition
for $ u_2 $). The "lowest energy eigenstate",
which is a bound
state, defines the critical temperature $ T^* $ below which the
interface state, Eq.(6), is unstable.
The corresponding wave function $ u_2(x) $ is nodeless and
decays exponentially in
the bulk region. For $ \delta > 0 $, as assumed above, the relative phase
$ \theta $ is $ \pm \pi/2 $ so that this state breaks time reversal
symmetry and is two-fold degenerate. It is not possible to obtain an
analytic solution of Eq.(8) in general. However, under the rather
restrictive condition $ b_2 = -\xi \coth(x_0/\xi) $ the instability
condition can be given analytically as $ \alpha_2(T^*) + u_1^2(T^*)
(\gamma-2\delta)/2=0 $ with the bound state wave function

\begin{equation}
u_2(x) = \frac{const.}{\cosh((x+x_0)/\xi)}.
\end{equation}

\noindent
Our analysis demonstrates that under certain conditions the interface
of an unconventional superconductor can give rise to a locally
$ {\cal T} $-violating state (see also \cite{SOU}). Furthermore,
$ T^* $ ($ < T_{c1} $) defined above is the
temperature at which a continuous phase transition from a
$ {\cal T} $-conserving ($ T > T^* $) to a $ {\cal T} $-violating
state ($ T < T^* $) occurs. It is, however,
not our aim to discuss here a possible microscopic basis for our GL-model.
Rather we are interested in some of the consequences of a
$ {\cal T} $-violating
superconducting phase.

Let us now study the phenomena which
occur at a Josephson junction between two superconductors $A$ and $B$
if $ {\cal T} $-violation is present. The
following discussion does not depend on whether the $ {\cal T} $-violation
is a bulk or, as discussed above, an interface (junction) phenomenon.
Because we have two complex order parameters at the interface,
the Josephson phase-current relation consists of four terms

\begin{equation}
J = \sum_{i,j=1,2} J_{ij} \sin(\phi_{iB} - \phi_{jA} )
\end{equation}

\noindent
where $J_{ij}$ are real constants whose sign and magnitude
depend on the grain orientation and
order parameter magnitude at the interface:
$ J_{ij} \propto |\eta_{iB} |
 |\eta_{jA}| \chi_i({\bf n}_B) \chi_j({\bf n}_B)
$, $ {\bf n}_{A,B} $ is the junction normal vector on either side
and, typically, $ \chi_1({\bf n}) = n^2_x - n^2_y $ and
$ \chi_2({\bf n}) = n_x n_y $.  We assume that the current
through the interface vanishes, because due to screening
effects (on a length scale $\lambda_J$), such currents can only flow
near the boundary of the interface or near a vortex.  Furthermore, we
assume that the couplings $J_{ij} $ are sufficiently weak so that the relative
phase between $ \eta_1 $ and $ \eta_2 $
is not affected, i.e.: $\phi_{1A}-\phi_{2A}=\phi_{1B}-\phi_{2B}= \pm \pi/2 $
in the $ {\cal T} $-violating state. This simplification is not important
for any of our later conclusions and a more complete discussion will be
given elsewhere.

The latter assumption allows us to minimize the junction energy,
$ E = -(\Phi_0/2 \pi c) \sum_{i,j} J_{ij} \cos(\phi_{iB} - \phi_{jA}) $,
by choosing the phases such that $J=0$.  We obtain

\begin{equation}
\Delta\phi_a = \phi_{iB}-\phi_{iA}= \pm\tan^{-1}
\left( \frac{J_{12}-J_{21}}{J_{11}+J_{22}} \right)
\end{equation}

\noindent
for a junction $ a $ with all $J_{ij} > 0$, and

\begin{equation}
\Delta\phi_b = \phi_{iB}-\phi_{iA}= \pi\pm\tan^{-1}
\left( \frac{J_{12}-J_{21}}{J_{11}+J_{22}} \right)
\end{equation}

\noindent
for a junction $ b $ with $J_{12} $, $J_{12} > 0$ and $J_{21}$, $J_{22} < 0$.

We consider now the situation where these two types of junctions, $ a $ and
$ b $, intersect (forming a grain boundary corner).
Such a corner is accompanied with phase winding or a vortex, because,
in general, $ \Delta \phi_a \neq \Delta \phi_b $. For the calculation of
the magnetic flux of this vortex we notice that the supercurrent is given
by the expression

\begin{equation}
\frac{2 \Phi_0}{c} {\bf j} = \sum_{i=1,2} K_i u^2_i (\nabla \varphi -
\frac{2 \pi}{\Phi_0} {\bf A} + \nabla \phi_i)
\end{equation}

\noindent
with $ \eta_j= u_j e^{i (\phi_j + \varphi)} $, j=1,2, and $ \varphi $ a
phase of the order parameter continuous even at the junction and
$ 0 \leq \phi_j \leq 2 \pi $ . We choose
a path $ C $ encircling the corner at a distance
far enough so that $ {\bf j}=0 $
on $ C $. We denote the segments of $C$ in superconductors $A$ and $B$
 by $C_A$ and $C_B$ respectively.
Using $ \phi_1 = \phi_2 \pm \pi/2 $ the circular
integral of Eq.(12) on $ C $ leads to the flux

\begin{equation}
\frac{\Phi}{\Phi_0} = n + (\int_{C_A} d{\bf s} + \int_{C_B} d{\bf s}) \cdot
\nabla \phi_1 = n + \frac{\Delta \phi_a - \Delta \phi_b}{2 \pi}
\end{equation}

\noindent
where $ n $ is the integer winding number of $ \varphi $.
Obviously, the flux at the corner can have any fraction of
$ \Phi_0 $ and is determined only by the properties of the junctions.
On the other hand, it is easy to see from our discussion that in the
case of a $ {\cal T} $-conserving superconducting state the only
fractional vortex is that with half a flux quantum $ \Phi_0$
($ \Phi = \Phi_0(n +1/2) $) \cite{VBG,AJM}. The field distribution
of such vortices would extend along the junction on a length scale
$ \lambda_J $ while penetrating the bulk only by the London
penetration depth $ \lambda \ll \lambda_J $.

The twofold degeneracy of the $ {\cal T} $-violating interface state implies
the existence of domains and domain walls. There is
a phase winding and flux associated with the intersection of a
domain wall and a grain boundary, because the phase
jump $ \Delta \phi $ at the junction is different on the right and left
hand side of a domain wall.
Following above scheme, a domain wall on junction $ a $ contains a flux
$ \Phi/\Phi_0 = n + \Delta \phi_a/ \pi $.
These vortices are similar to the fractional
domain wall vortices analyzed in Ref.\cite{SRU}. They are not
connected with corners, but can essentially by located anywhere on
a grain boundary. Hence, we may conclude that our model can account
for both fractional
vortices at the corners and along the edges of the triangle as observed
by Kirtley et al. \cite{KIR}.

Let us make several remarks.  The central point of this work is that
{\em any} superconducting state with fractional vortices containing
other than
($n+1/2$) flux quanta
violates time reversal symmetry.  Fractional vortices are
not specific to the $d_{x^2-y^2}+i\epsilon
d_{xy}$ state, although in the interest of
simplicity we restricted ourselves to this order parameter in our model
calculations.  It should also be
noted that on the basis of the experiment by Kirtley and co-workers
alone, the specific
form of ${\cal T}$-violation cannot be deduced.  Our conclusion that time
reversal symmetry breaking has been seen can only be wrong if the
assumption of existence of the fractional vortices is wrong.
There are several experiments which would add
considerably to our understanding of this matter.  One such experiment
would be to look for the critical temperature $T^*$ at which
${\cal T}$-violation occurs. According to our discussion there would have to
be a second phase transition below the onset of superconductivity, although
this could very likely be a grain boundary phenomenon only.
Above this temperature $ T^* $ there can be no fractional vortices
apart from those with $ \Phi= \pm \Phi_0/2 $.
It would also be interesting to look for fractional vortices in
different materials such as HgBa$_2$Ca$_2$Cu$_3$O$_8$ or
Bi$_2$Sr$_2$CaCu$_2$O$_8$, perhaps utilizing different
geometries which might better isolate the grain boundary corners.
The existence of a complex order parameter at the interface
may also shed new light on the interpretation of Josephson junction
experiments such as those of Chaudhari {\em et al}.\cite{CHA} and
Sun {\em et al.}\cite{SUN}, beyond the analysis given recently by Millis
\cite{AJM}.

In this Letter we have shown that (1) the existence of vortices enclosing a
fraction of a flux quantum requires the breaking of time reversal symmetry,
and (2) that the converse is also true.
We argue that this has been observed at grain boundaries in
YBa$_2$Cu$_3$O$_7$.
Further experiments are needed to deduce the nature and extent of
${\cal T}$-violation in high temperature superconductors.

We are grateful to J. Kirtley, P.A. Lee, A. Millis, T.M. Rice and
A. Furusaki for helpful discussions. This research was supported NSF Grant
No. DMR-88-16217.  D. B. B.
acknowledges a fellowship from the National Science Foundation and M.S.
a fellowship from Swiss Nationalfonds.

\end{document}